\documentclass[runningheads]{llncs}
\usepackage{graphicx}
\usepackage{array}
\usepackage{cite} % sort citations numerically
\usepackage{enumitem}

\begin{document}

%\title{A Platform for Cybersecurity Awareness Training with an Automated Coach\thanks{Supported by organization x.}}
\title{Cybersecurity Awareness Platform with Virtual Coach and Automated Challenge Assessment}

\titlerunning{Cybersecurity Awareness with Virtual Coach and Automated Assessment}

\author{
Tiago Gasiba\inst{1,2}\orcidID{0000-0003-1462-6701} \and
Ulrike Lechner\inst{2}\orcidID{0000-0002-4286-3184} \and
Maria Pinto-Albuquerque\inst{3}\orcidID{0000-0002-2725-7629} \and
Anmoal Porwal\inst{1}\orcidID{0000-0002-2926-9797}
}

\authorrunning{T. Gasiba et al.}

\institute{
Siemens AG, Munich, Germany \\
\email{\{tiago.gasiba,anmoal.porwal\}@siemens.com}
\and
Universität der Bundeswehr München, Munich, Germany
\email{\{ulrike.lechner,tiago.gasiba\}@unibw.de}
\and
Instituto Universitário de Lisboa (ISCTE-IUL), ISTAR-IUL, Lisboa, Portugal\\
\email{maria.albuquerque@iscte-iul.pt}
}

\maketitle

\begin{abstract}
Over the last years, the number of cyber-attacks on industrial control systems has been steadily increasing.
Among several factors, proper software development plays a vital role in keeping these systems secure.
To achieve secure software, developers need to be aware of secure coding guidelines and secure coding best practices.
This work presents a platform geared towards software developers in the industry that aims to increase awareness of secure software development.
The authors also introduce an interactive game component, a virtual coach, which implements a simple artificial intelligence engine based on the laddering technique for interviews.
Through a survey, a preliminary evaluation of the implemented artifact with real-world players (from academia and industry) shows a positive acceptance of the developed platform.
Furthermore, the players agree that the platform is adequate for training their secure coding skills.
The impact of our work is to introduce a new automatic challenge evaluation method together with a virtual coach to improve existing cybersecurity awareness training programs. These training workshops can be easily held remotely or off-line.

\end{abstract}

% ---- Core of the Paper ----
\section{Introduction}
Errors and vulnerabilities in software development, if not solved early, can end up in a final product.
These problems can result in serious consequences for the customer and the company that produced the software.
This work aims to improve the situation through a serious game to raise awareness on secure coding and software development best practices of software developer -- thus addressing the issues at early stages in software development, i.e., when it is being written.

In the next sub-sections, we present the problem at hand in more detail. We give a brief overview of standardization bodies, industry-led efforts, and academic efforts that were started to address the current situation.
Finally, we describe our proposed methodology and our contributions to scientific knowledge.

\subsection{The Need for Secure Coding Awareness}
The number of security advisories issued per year by the Industrial Control System - Computer Emergency Response Team (ICS-CERT) has been steadily increasing.
While before 2014 the number of advisories per year was less than 100, from 2017 to 2019 more than 200 advisories have been issued per year.
These facts correlate well with the observed increase in the number and sophistication of cyber-attacks to industrial control systems (ICS).

The ransomware WannyCry, released by the "The Shadow Brocker" hacker group in 2017, which exploits a vulnerability in the Server Message Block (SMB) protocol, dubbed EternalBlue, has affected numerous industrial control systems.
It has caused a financial impact exceeding 4 billion USD in more than 140 countries.
The vulnerability exploited by EternalBlue is a buffer overflow caused by an integer overflow; exploitation of buffer overflows is not new - this is known since the late '70s.

While not everything (e.g., attacks and vulnerabilities) can be traced back directly to a specific software vulnerability, an increasing number of such vulnerabilities (i.e., related with secure coding) have also been observed.
Software security and secure software development play a fundamental role in industrial cybersecurity, particularly in critical infrastructures.
According to a recent survey with more than 4000 software developers~\cite{gitlab_2019}, \textit{"less than half of developers can spot security holes"}~\cite{schneier_2019_sw_devel}.
This lack of awareness causes a severe issue in terms of cybersecurity of industrial control systems and critical infrastructures.
The present work focuses on C and C++ programming languages.
This is motivated by a recent study by Whitehat~\cite{WhiteSource2019}, which has shown that C and C++ are among the most used programming languages for industrial environments, but they are also among the most vulnerable in terms of cybersecurity vulnerabilities.
This study also implies that the majority of vulnerabilities are created in these programming languages.

\subsection{Standards, Industry, and Academic Efforts}
In recognition of the importance of secure products and a consequence of the current move towards digitalization and higher connectivity, several large industrial players have joined together and committed to a document called the charter of trust \cite{siemens02_charter}.
The charter of trust outlines ten fundamental principles that the partners vow to obey to address the issues inherent with cybersecurity.
ICS relevant standards such as IEC 62443-4-1~\cite{2018_62443_4_1} or ISO 27001~\cite{2013_27001} mandate the implementation of secure software development life-cycle processes and awareness training.
These standards address security from a high-level perspective and are not specific enough about recommendations, policies, and best practices to be followed in software development.
Towards this goal, an industry-led effort was created, the Software Assurance Forum for Excellence in Code (SAFECode), with the aim of {\it identifying and promoting best practices for developing and delivering more secure and reliable software, hardware and services}.

Serious Games designed to train developers and raise their awareness for cybersecurity and secure coding is our approach to ameliorate the situation, and other approaches are ~\cite{gasiba_re19,2018_rieb_IT_sicherheit,2017_rieb_gamified_approach,Maria2019}. We designed a game to raise awareness for cybersecurity among programmers and for secure coding guidelines and secure coding best practices. Our approach is an adoption of the popular format of Capture-the-Flag.
CTF is a serious game genre in the domain of cybersecurity popular in the penetration-test community as a means to practice offensive skills.
In this kind of game, the game participants aim to gather the highest amount of awarded by solving cybersecurity challenges, e.g., breaking into systems.
In~\cite{gasiba_re19}, Gasiba et al. study the requirements that a game designer should follow to target the game to software developers in the industry.
In a further work~\cite{gasiba2020_challenge_types}, the authors provide six concrete and different challenge types to be used in this kind of CTF event.
One of these is the "code entry" challenge type, where the proposed idea is that player interacts through a web interface with a backend by modifying vulnerable code until all the coding guidelines are fulfilled, thus solving the challenge.

\subsection{Automatic Challenge Evaluation}

This paper extends the previous work, particularly the "code entry" challenge type, by describing the architecture of a platform, which the authors call Sifu, that was constructed to implement the game backend.
The goal of this platform is to: 1) automatically analyze the solution submitted by the participant to the backend, 2) determine if this solution contains vulnerabilities and fulfills the required functionality, 3) generate hints to the player if the solution does not achieve a pre-determined goal and finally 4) provide a flag (i.e., a unique code) which the player can use to gather points in the game.
The correctness of the solution depends on it following established secure coding guidelines and secure programming best practices.

The generated hints are provided by a virtual coach, which assists the player in solving the challenge.
These hints are created using a simple artificial intelligence (AI) engine that provides automatic pre-programmed interactions with the player when the submitted solution fails to meet the secure coding criteria.
These hints generated by the AI Engine (i.e., the virtual coach) assist the player in solving the challenge in a playful way and help lower the frustration, increase the fun, and improve the learning effect during gameplay.

The core of the present work is to describe the virtual coach platform.
Nevertheless, to validate its suitability as a means to raise secure coding awareness, a small survey was performed with real players.
Our preliminary results show that the participants have fun using the platform and also find it adequate for learning secure coding guidelines and secure software development best practices.

\subsection{Contributions of this work}

This work seeks to provide the following impact in the research community:
\begin{itemize}
    \item introduce a novel method to automatically analyze player code submission in terms of secure coding guidelines and software development best practices,
    \item introduce a virtual coach based on the laddering interview AI technique, and
    \item provide a preliminary analysis of the suitability of the proposed architecture in terms of adequacy to raise secure coding awareness of software developers.
\end{itemize}

Although we intend to use the Sifu platform in a CTF environment, it can also be used stand-alone in remote and offline training scenarios.
This can be especially important if the players are spread over a large geographic area or have inherent restrictions on a face-to-face workshop.

\subsection{Paper Outline}
This work is organized as follows.
In section~\ref{sec:related_workd} we present previous related scientific work.
Section~\ref{sec:sifu_platform} presents details on the architecture and implementation of the Sifu platform.
This section also introduces the virtual coach and gives details on the implemented artificial intelligence algorithm.
In section~\ref{sec:results}, preliminary results from a short survey to 15 participants in a pilot are presented.
Finally, section~\ref{sec:conclusions} presents the conclusions and further work.
\section{Related Work}
\label{sec:related_workd}

Playing cybersecurity games is gaining more and more attention in the research community.
In~\cite{Maria2019}, Frey et al. show both the potential impact that playing cybersecurity games can have on the participants and also show the importance of playing games as means of cybersecurity awareness.
They conclude that cybersecurity games can be useful to {\it build a common understanding of security issues}.

A serious game~\cite{2016_Doerner_Serious_Games} is a game that is designed with a primary goal and purpose other than entertainment.
Typically these games are developed to address a specific need such as learning or improving a skill.
A Capture-the-Flag (CTF) game is one possible instance of a serious game.
Votipka et al.~\cite{votipka2018toward} argue in their work that CTF events can be used as a means to improve security software development.
In particular, their work shows that the participants of such events experience positive effects on improving their security mindset.
Davis et al., in~\cite{Davis2014}, discuss the benefits of CTF for software developers.
In their work, they argue that CTFs can be used to teach computer security and conclude that playing CTFs is a fun and engaging activity.

In their work, Graziotin et al.~\cite{2018_Graziotin_Happy_Developers} argue that \textit{happy developers are better coders}.
They show that developers that are happy at work tend to be more focused, adhering to software development processes, and following best practices.
This improvement in software development leads to the conclusion that happy developers can produce higher quality and more secure code than unhappy developers.
The authors believe that CTF events since they are experienced as fun events, can foster higher code quality and adherence to secure development principles.

However, CTF events need to be properly designed to achieve this goal.
Gasiba~et~al., in~\cite{gasiba_re19}, perform requirements elicitation employing systematic literature review, interview of security experts, and also CTF participants from industry.
Their work details the requirements for CTF events to raise secure coding awareness of software developers in the industry.
In particular, they conclude that CTF challenges for software developers should focus on the defensive perspective instead of offensive.

In their work, Simões et. al~\cite{simoes_icpec2020} present several programming exercises for teaching software programming in academia.
Their design includes nine exercises that can be presented to students to foster student motivation and engagement in academic classes and increase learning outcomes.
Their approach uses gamification and automatic assessment tools.
However, their work focus on the correct solution (implementation) of the programming exercise and not on the secure programming and security best practices aspects.

Gasiba et. al~\cite{gasiba2020_challenge_types} propose, in a similar work, six different challenge types.
These challenges, which are also a form of programming exercises, are executed in the context of a serious game of the type CTF and target software developers in the industry.
One of the challenge types is a so-called code-entry challenge, where the CTF participant is given a project (e.g., in C or C++) that contains software vulnerabilities.
The challenge aims to have the participants fix the security vulnerabilities by applying secure coding guidelines and software development best practices.
In this previous work, the challenge type was only derived conceptually and lacked implementation and practical evaluation aspects.

Vasconcelos et. al~\cite{vasconcelos_icpec2020} have recently shown a method to evaluate programming challenges automatically.
In their work, the authors use Haskell and the QuickCheck library to perform automated functional unit tests of challenges submitted by students.
Their goal is to evaluate if the solutions presented by the students comply with the programming challenge in terms of desired functionality.
One of the main limitations of this work is that the code to be tested should be free from side effects.
The authors also focus on functional testing of single functions and do not address the topic of cybersecurity.

In~\cite{dobrovsky2016approach,brisson2012artificial}, Dobrovsky et al. describe an interactive reinforcement learning framework for serious games with complex environments where a non-player character is modeled using human guidance.
They argue that interactive reinforcement learning can be used to improve learning and the quality of learning.
However, their work aims to train an algorithm better to recreate human behavior by means of machine learning techniques.
In our work, we aim at training humans to write better and more secure code.
Due to this fact, machine learning techniques are not applicable.
Nonetheless, we draw inspiration from the conceptual framework, which we adapt to our scenario.

Rietz et al.~\cite{rietz2019ladderbot}, show how to apply the principles of the laddering interview technique for requirements elicitation.
The laddering technique consists of issuing a series of questions that are based on previous system states (i.e., previous answers and previous questions).
The questions generated are refined versions of previously issued questions as if the participant is climbing up a ladder containing more specific questions.
Although this previous work applies in the field of requirements elicitation and does not focus on cybersecurity, the laddering technique principle can be adapted to a step-wise hint system.

In the present work, we also make use of the concept of awareness or IT-security awareness as defined by Haensch et al. in~\cite{2014_Benenson_Defining_Security_Awareness}, in order to evaluate our artifact.
In their work, they define awareness as having the following three dimensions: perception, protection, and behavior.
The perception dimension is related to the knowledge of existing software vulnerabilities.
The protection dimension is related to knowing the existing mechanisms (best practices) that avoid software vulnerabilities.
Finally, the behavior dimension relates to the knowledge and intention to write secure code.
We collect data from participants based on the three dimensions of awareness through a small survey.
We use best practices in the design, collection, and processing of survey information given by Grooves et al.~\cite{Groves2009}.

\section{Sifu Platform}
\label{sec:sifu_platform}
In following sub-sections we present the research problem in terms of research questions and present a possible solution.
Additionally, we describe the setup of a small survey that was performed to evaluate our result.

%%%%%%%%%%%%%%%%%%%%%%%%%%%%%%%%%%%%%%%%%%%%%%%%%%%%%%%%%%%%%%%%
%%%%%%%%%%%%%%%%%%%%%%%%%%%%%%%%%%%%%%%%%%%%%%%%%%%%%%%%%%%%%%%%
%%%%%%%%%%%%%%%%%%%%%%%%%%%%%%%%%%%%%%%%%%%%%%%%%%%%%%%%%%%%%%%%
\subsection{Problem Statement}
In~\cite{gasiba2020_challenge_types}, the authors present a type of challenge for CTFs in the industry, which is called code-entry challenge (CEC).
The main idea of this type of challenge is for the Player to be given a software development project that contains code that does not follow secure coding guidelines (SCG) and secure software development best practices (BP) and contains security vulnerabilities.
In this work, we target specifically ICS by using SCG and BP, which are specific for this field.
The task of the Player is to fix the vulnerabilities and to follow SCG and BP.
The Player should do this so that the original intended functionality is still fulfilled in the new version of the code.
The present work aims to solve these requirements by means of a platform that performs an automatic evaluation of the code submitted by the participant and guides the participant towards the final solution.
Considering these requirements, the following research questions are then raised:

\begin{itemize}[leftmargin=+.45in]
    \item[ {\bf RQ1:}] how to automatically assess the challenges in terms of SCG and BP?
    \item[ {\bf RQ2:}] how to aid the software developer when solving the challenges?
\end{itemize}

This work proposes to address RQ1 through a specialized architecture to automatically assess the level of compliance to SCG and BP by combining several state-of-the-art security testing frameworks, namely Static Application Security Testing (SAST), Dynamic Application Security Testing (DAST), and Runtime Application Security Protection (RASP).
The functional correctness of the provided solution by the Player is evaluated using state-of-the-art Unit Testing (UT).
To address RQ2, the authors propose to combine the output of the security testing tools with an AI algorithm to generate hints based on the laddering technique, thus implementing a virtual coach.
The task of the virtual coach is to lower the frustration of the participant during gameplay and to aid in the participant to improve the code.

The proposed solution herein described makes a contribution towards answering these research questions.
To validate the assumption of the suitability of our proposal as a means to address the research questions, a small survey was conducted.

%%%%%%%%%%%%%%%%%%%%%%%%%%%%%%%%%%%%%%%%%%%%%%%%%%%%%%%%%%%%%%%%
%%%%%%%%%%%%%%%%%%%%%%%%%%%%%%%%%%%%%%%%%%%%%%%%%%%%%%%%%%%%%%%%
%%%%%%%%%%%%%%%%%%%%%%%%%%%%%%%%%%%%%%%%%%%%%%%%%%%%%%%%%%%%%%%%
\subsection{Code-entry challenge platform architecture}
\begin{figure}[ht]
    \centering
    \includegraphics[width=.7\columnwidth]{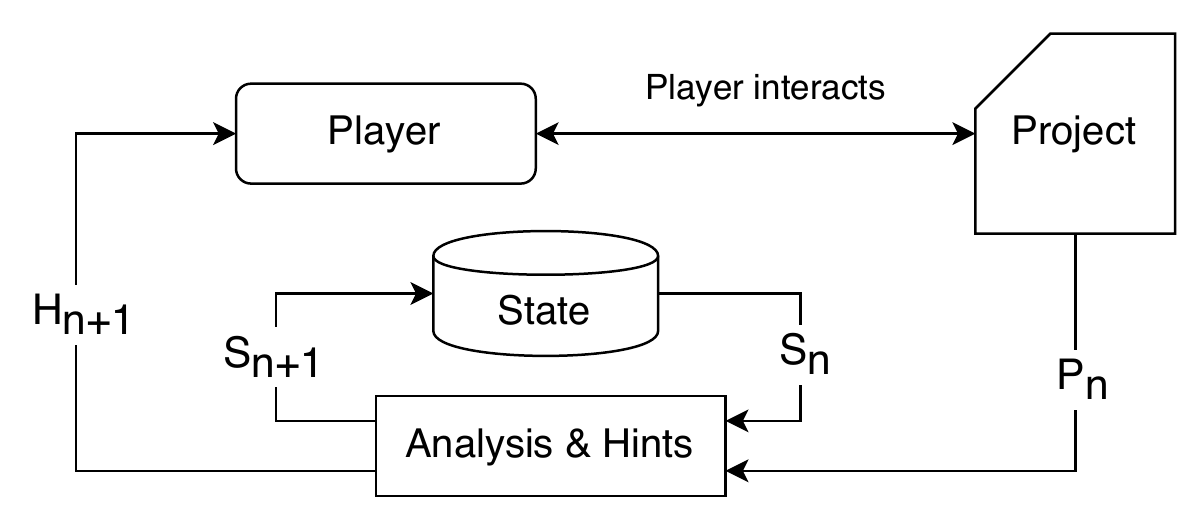}
    \caption{Conceptual game overview: interaction and components}
    \label{fig:game_overview}
\end{figure}

Figure~\ref{fig:game_overview} shows the top-level view of the Sifu architecture.
In this figure, the "Player" represents the game participant (a human) and the "Project" represents a software project that contains vulnerabilities to be fixed by the Player.
The "Analysis \& Hints" (AH) component performs the core functionality: 1) evaluates the submitted code (Project) in terms SCG and BP, 2) indicates if the challenge is solved or not and, if not solved, 3) generates hints to send back to the participant. 
The "State" component stores previous interactions and generated hints.
During gameplay, the Player reads the Project and modifies the code by interacting with a web editor interface.
When the changes in the code are done, the Player submits the code to the AH component for analysis.
 
\begin{figure}[ht]
    \centering
    \includegraphics[width=.95\columnwidth]{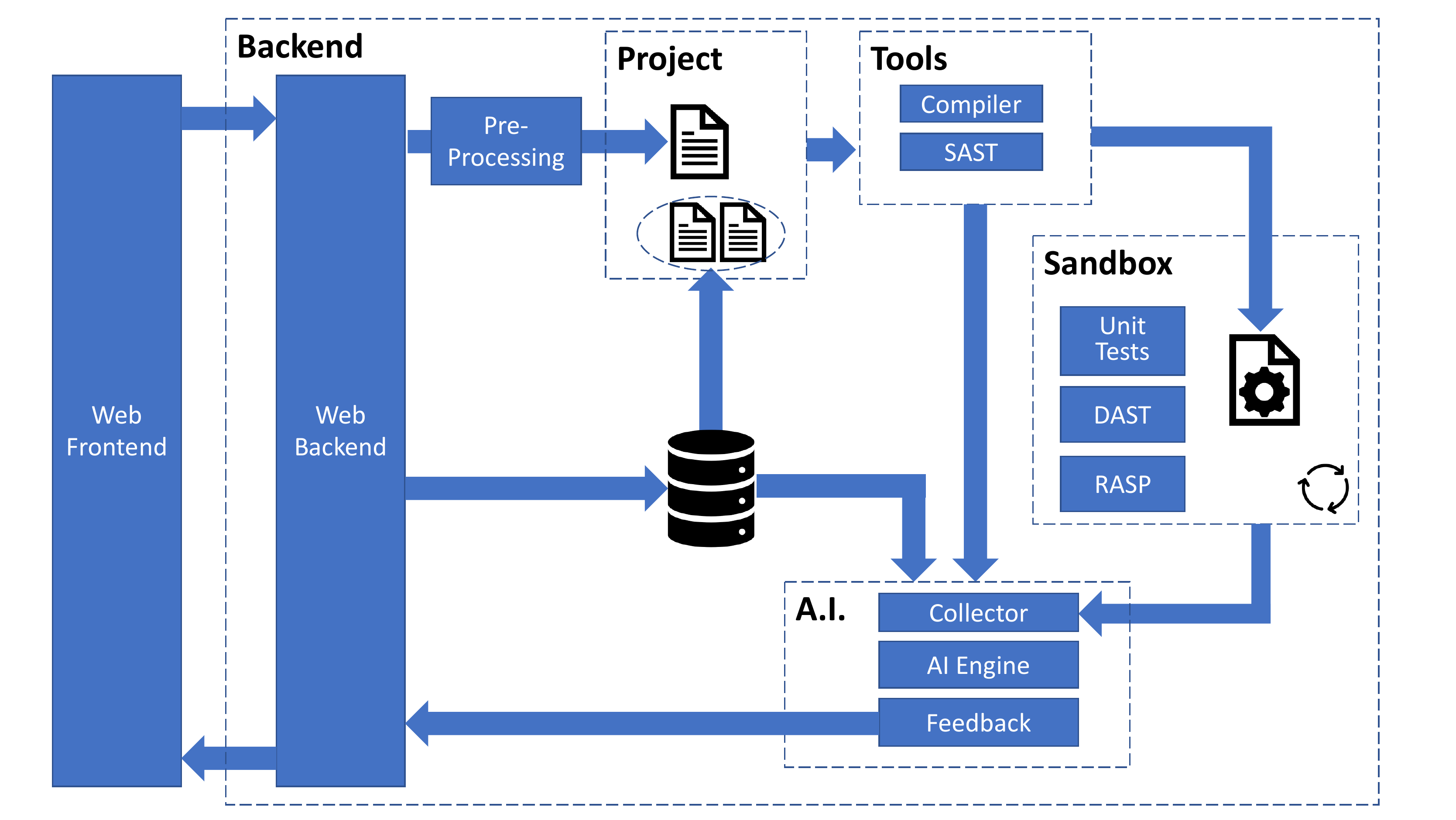}
    \caption{Detailed architecture: the Sifu Platform}
    \label{fig:sifu_architecture}
\end{figure}

A possible realization of the conceptual architecture is shown in figure~\ref{fig:sifu_architecture}.
Interaction takes place between the Player and a web interface, which connects to a web backend.
The web backend is responsible for triggering the automated security assessment, collecting the answer from the AI engine, and sending the answer back to the participant.
To realize this, the Project submitted by the participant is first saved into a temporary folder after a pre-processing step (e.g. to inject code necessary for unit tests).
After the addition of auxiliary files (e.g. C/C++ include files) to the temporary project directory, the Project is compiled, and a functional test and security assessment is performed.
All these results are then made available to an AI engine, which determines if the challenge is solved and generates hints.
This feedback is collected by the web backend and stored in an internal database and forwarded as the answer back to the participant's web browser.

%%%%%%%%%%%%%%%%%%%%%%%%%%%%%%%%%%%%%%%%%%%%%%%%%%%%%%%%%%%%%%%%
%%%%%%%%%%%%%%%%%%%%%%%%%%%%%%%%%%%%%%%%%%%%%%%%%%%%%%%%%%%%%%%%
%%%%%%%%%%%%%%%%%%%%%%%%%%%%%%%%%%%%%%%%%%%%%%%%%%%%%%%%%%%%%%%%
\vspace{-1cm}
\subsubsection{Automatic Security Assessment}
The security assessment which is performed to the Project is composed of the following steps: 1) Compilation, 2) Static Application Security Testing, 3) Unit Testing, 4) Dynamic Application Security Testing, and 5) Runtime Application Security Testing.
In step 1, the Project is compiled; if there are compilation errors, these are reported to the AI component, and no further analysis takes place.
Step 2 performs static code analysis. Note that in this step, the code does not need to be executed.
Since the steps 3, 4 and 5 involve executing untrusted (and potentially dangerous) code, these are performed in a time-limited sandbox.
The sandbox is very restrictive, e.g., it only contains the project executable and drops security-relevant capabilities (e.g., debugging and network connections are not allowed).
Additionally, the executable is only allowed to run for a certain amount of time inside the sandbox.
If this time is exceeded, the process will be automatically terminated.
This avoids denial-of-service attacks by means of high CPU usage.
Two types of Unit tests are executed: 1) functional testing - in order to guarantee that the provided code is working as intended (e.g., in the challenge description), and 2) security testing - in order to guarantee that typical vulnerabilities are not present in the code (e.g., buffer overflow).
Security testing is done using self-developed tests and also using state-of-the-art fuzzing tools.
Steps 4 and 5 perform several dynamic security tests.
Table~\ref{table:sec_tools} lists the tools that the authors have used in each of these components. In this table, the open-source components used in the Sifu platform are marked with "OS".
\begin{table}[ht]
  \centering
  \renewcommand{\arraystretch}{1.1}
  \scriptsize
  \caption{Security Assessment Tools}
  \label{table:sec_tools}
  \begin{tabular}{|m{2cm}|p{8.5cm}|}
  \hline
  {\bf ~~Component} & {\bf ~~~~~~~~~~~~~~~~~~~~~~~~~~~~~~~~Tools}  \\
  \hline
  \hline
  ~~~~~Compiler & GCC v10.1 (OS), Clang 9.0.0 (OS) \\
  \hline
  ~~~~~~~SAST & SonarQube, Pc Lint, cppchecker (OS), fbinfer (OS), semgrep (OS) \\
  \hline
  ~~~~~~~DAST & Valgrind (OS), Helgrind (OS)\\
  \hline
  ~~~~~~~RASP & Address Sanitizer (OS), Leak Sanitizer (OS), Thread Sanitizer (OS) \\
  \hline
  ~~~~~Unit Test & ATF (OS), Kyua (OS), AFL (OS)\\
  \hline
  \end{tabular}
\end{table}

%%%%%%%%%%%%%%%%%%%%%%%%%%%%%%%%%%%%%%%%%%%%%%%%%%%%%%%%%%%%%%%%
%%%%%%%%%%%%%%%%%%%%%%%%%%%%%%%%%%%%%%%%%%%%%%%%%%%%%%%%%%%%%%%%
%%%%%%%%%%%%%%%%%%%%%%%%%%%%%%%%%%%%%%%%%%%%%%%%%%%%%%%%%%%%%%%%
\subsubsection{Virtual Coach with AI Technique}
\label{lab:laddering_techinque}
\begin{figure}[ht]
    \centering
    \includegraphics[width=.80\columnwidth]{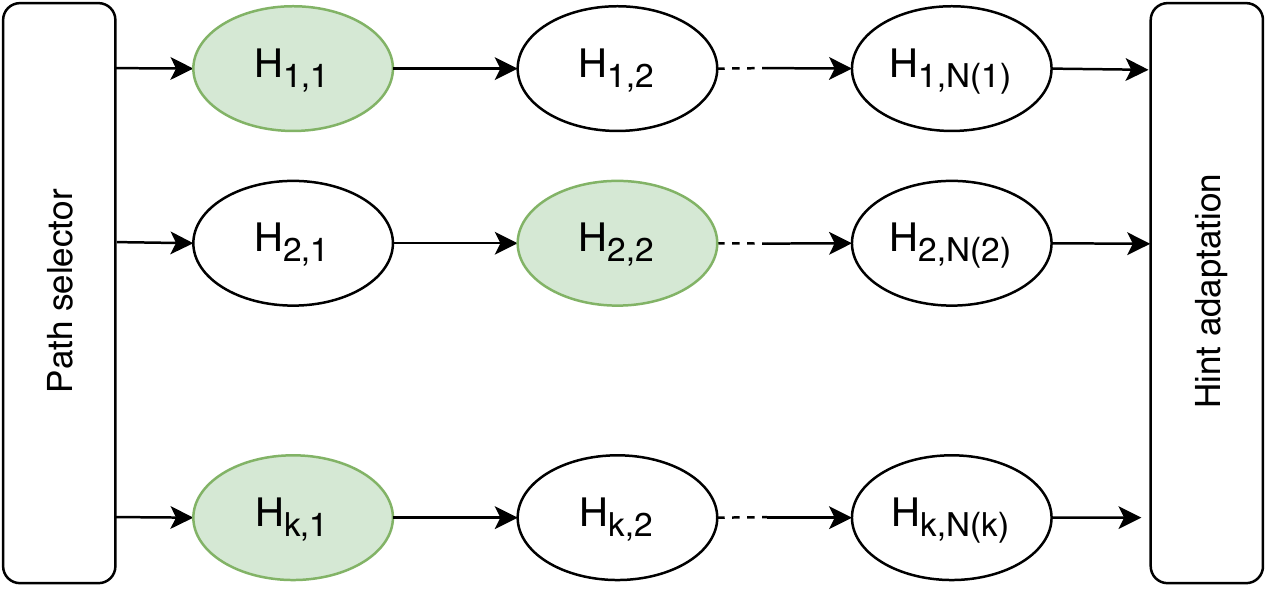}
    \caption{Laddering technique to generate hints}
    \label{fig:laddering}
\end{figure}

The AI component shown in figure~\ref{fig:sifu_architecture} collects the results of the previous analysis steps, runs an AI engine based on the laddering technique, and generates the feedback to be sent back to the participant.
Figure~\ref{fig:laddering} shows the implementation of the AI engine using the laddering technique.

As previously detailed, the automated assessment tools perform several tests that are used to determine the existing software vulnerabilities present in the Project.
These are collected in textual form (e.g., JSON and XML) and normalized to be processed by the AI engine.
The two most essential test results from the security assessment components are related to compilation errors (e.g., syntax errors) and functional unit testing.
The participant's solution will be rejected if the code does not compile or is not working (functioning) as intended.
When both these tests pass, the artificial engine uses the security tests, SAST, DAST, and RASP tools to generate hints to send to the participant.

A combination of findings from these tools forms a vulnerability.
These findings and vulnerabilities are then mapped to SCG and BP.
In figure~\ref{fig:laddering}, each horizontal path (ith row) corresponds to a ladder and also to a specific combination of vulnerabilities or static events found in the source code.
Each path is also assigned a priority $p(i)$ based on the criticality of the SCG and vulnerabilities.
These priorities are assigned according to the ranking of secure coding guidelines, as presented in Gasiba et al. (see~\cite{gasiba2020_ranking_scg}).
Higher-ranked secure coding guidelines are given higher priorities, and lower-ranked secure coding guidelines are given lower priorities.
The AI engine to selects the corresponding path (corresponding to one ladder) which based on the finding with the highest rank.

%In this stage, the possible hints are given by $H_{p,\lambda}$, where $p$ is the selected path.
The chosen hint $H_{n+1}$ depends on the ladder and on the previous hint level sent to the participant on the ladder, as given by the system state.
If there are no more hints in the ladder, no additional hint is sent to the Player.

Table~\ref{table:hint_levels} shows an example of hints provided by the virtual coach's AI engine corresponding to an "undefined behavior" path.
The lower level hints are generic and give background information for the participant.
The highest level hint contains exact information on how to solve the problem, thus revealing the solution.
\begin{table}[ht]
  \centering
  \renewcommand{\arraystretch}{1.1}
  \scriptsize
  \caption{Example of hint ladder with six levels}
  \label{table:hint_levels}
  \begin{tabular}{|m{1cm}|p{10cm}|}
  \hline
  {\bf ~Level} & {\bf ~~~~~~~~~~~~~~~~~~~~~~~~~~~~~~~~~~~~~~~~~Hint Text} \\
  \hline
  ~~~~1 & The following links contain information that might be helpful: $<$link$>$, $<$link$>$ \\
  \hline
  ~~~~2 & The compiler is free to optimize the compiled code assuming that there is no undefined behavior in the code \\
  \hline
  ~~~~3 & Look at the variable 'i' \\
  \hline
  ~~~~4 & Read carefully the following secure coding guideline: $<$link$>$ \\
  \hline
  ~~~~5 & The code accesses the variable "Values" - check carefully the bounds \\
  \hline
  ~~~~6 & Since undefined behavior is not allowed, and the variable "Values" must be indexed within the bounds, the check i$<$4 is removed by the compiler! \\
  \hline
  \end{tabular}
\end{table}

Finally, the Feedback component formats and enriches the selected hint by the AI Engine with project-specific information and sends it to the Web Back-End component to present to the Player.
To foster critical thinking, the authors have also implemented a hint back-off (i.e., no hint will be given to the Player who is brute-forcing the hint system).
This back-off system implements the following rule: 1) no hint is provided to the Player during 4 minutes after the backend has sent a hint to the Player, and 2) no hint is given until the number of code submissions since the previous hint sent to the Player by the backend is equal to 3 submissions.

Note that the feedback component not only fosters critical thinking by the Player, but can also be used to train the Player with the usage of static code analysis tools. However, further investigation of this aspect is needed in the future.

%%%%%%%%%%%%%%%%%%%%%%%%%%%%%%%%%%%%%%%%%%%%%%%%%%%%%%%%%%%%%%%%
%%%%%%%%%%%%%%%%%%%%%%%%%%%%%%%%%%%%%%%%%%%%%%%%%%%%%%%%%%%%%%%%
%%%%%%%%%%%%%%%%%%%%%%%%%%%%%%%%%%%%%%%%%%%%%%%%%%%%%%%%%%%%%%%%
\subsubsection{Real-World Artifact}
\begin{figure}[ht]
    \centering
    \includegraphics[width=.85\columnwidth]{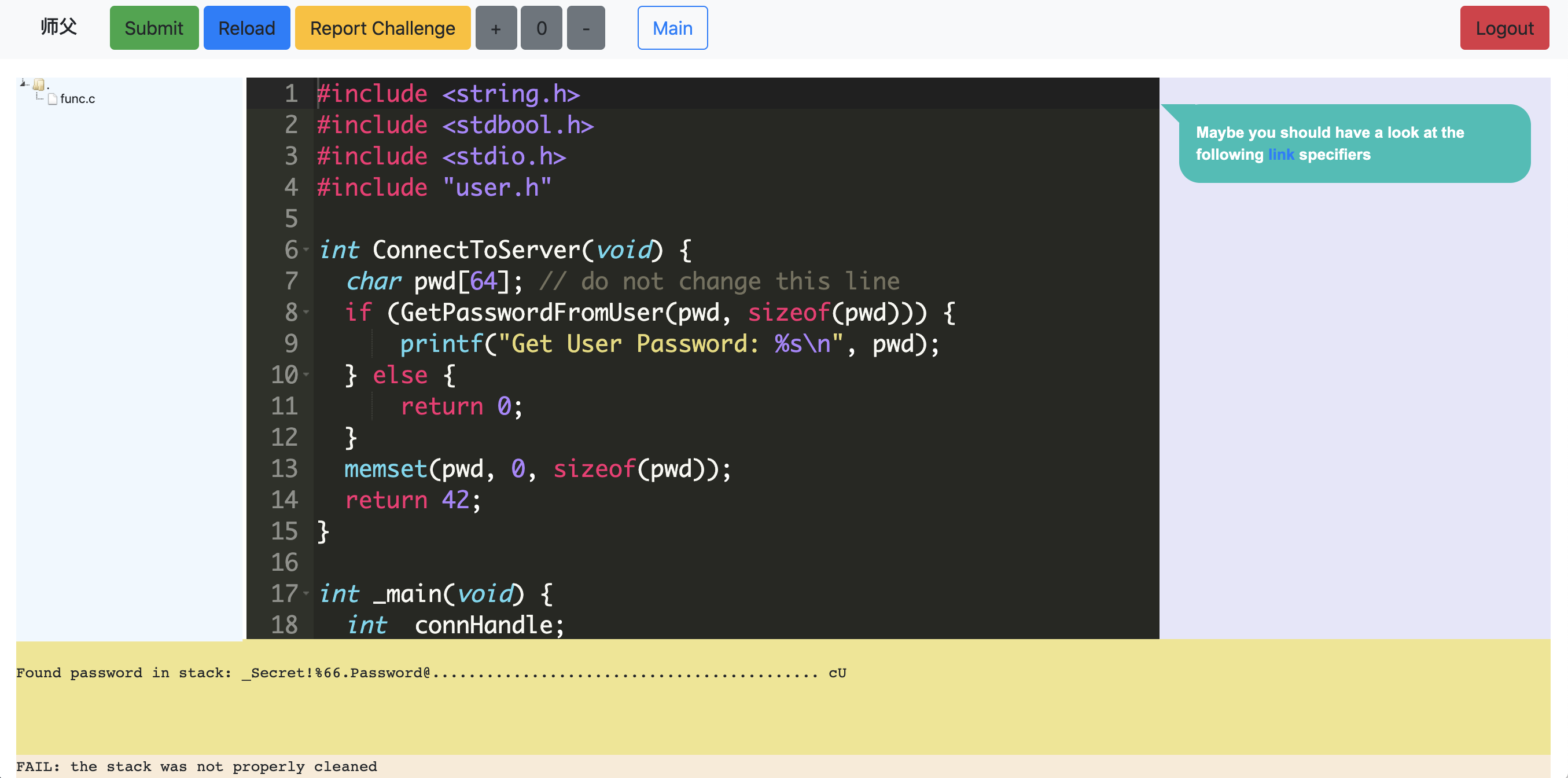}
    \caption{Sifu Web Interface}
    \label{fig:sifu_challenge}
\end{figure}

Figure~\ref{fig:sifu_challenge} shows the web interface of a real-world implementation of the Sifu platform.
The machine where the Sifu platform was deployed was an AWS instance of type T3.Medium (2 CPUs with 4Gb RAM and network connection up to 5Gb/s). In order to install the required tools, a hard-disk of 40Gb was selected. The Sifu platform itself is developed in Python 3.8 using Flask.

On the left, the Player can browse the Project and select a file to edit; the file editor is in the center, and on the right are the hints that the Player receives from the backend.
The upper part contains buttons which include the following functionality: {\it Submit} - to submit the Project for analysis, {\it Reload} - to reload the Project from scratch, {\it Report Challenge} - to report problems with the challenge to the developers.
Note that, when a player finishes a challenge successfully, it is taken to an additional page with discussions on the impact of the vulnerability and additional closing questions (e.g., on which secure coding guidelines have not been taken into consideration).

\subsubsection{Evaluation of real-world artifact}
The platform containing five different challenges was made available for experimentation to 15 participants in Germany in June 2020.
Participants' ages ranged between 20 and 50 years old, with an average of $28.3$.
The participants' background was: $7$ computer science students, $7$ professional software developers, and $1$ assistant professor.
Participants were allowed to try the platform for as long as they liked; this resulted in a range from 15 minutes to 45 minutes.
When successfully solving a challenge, the participants were asked (through the web interface) to rate the challenge based on the questions presented in table~\ref{table:chal_rating}.
Additionally, upon completing the experiment, when the participants were asked to fill out a small online survey.
The questions asked in this survey are presented in table~\ref{table:platform_questions}.
Both the challenge rating and the platform survey questions were based on a 5-point Likert scale.

\begin{table}[ht]
  \centering
  \renewcommand{\arraystretch}{1.1}
  \scriptsize
  \caption{Challenge rating questions}
  \label{table:chal_rating}
  \begin{tabular}{|m{2cm}|p{7.5cm}|}
  \hline
  {\bf ~~~~Number} & {\bf ~~~~~~~~~~~~~~~~~~~~~~~~~~~~~Question} \\
  \hline
  ~~~~~~~~~Q1 & Please give an overall rating to the challenge \\
  \hline
  ~~~~~~~~~Q2 & How well could you recognize the vulnerability in the code? \\
  \hline
  ~~~~~~~~~Q3 & How well can you fix this problem in production code? \\
  \hline
  \end{tabular}
\end{table}

\begin{table}[ht]
  \centering
  \renewcommand{\arraystretch}{1.1}
  \scriptsize
  \caption{Platform survey questions}
  \label{table:platform_questions}
  \begin{tabular}{|m{1.2cm}|p{7.5cm}|}
  \hline
  {\bf Number} & {\bf ~~~~~~~~~~~~~~~~~~Feedback Question}  \\
  \hline
  ~~~~~F1 & My overall experience with the platform was positive \\
  \hline
  ~~~~~F2 & The Sifu platform helps me to improve my secure coding skills  \\
  \hline
  ~~~~~F3 & Solving challenges in the Sifu platform helps me in recognizing vulnerable code\\
  \hline
  ~~~~~F4 & Solving challenges in the Sifu platform helps me in understanding consequences of exploiting vulnerable code \\
  \hline
  ~~~~~F5 & Solving challenges in the Sifu platform makes me overall happy \\
  \hline
  ~~~~~F6 & Challenges in the Sifu platform help me to practice secure coding guidelines \\
  \hline
  ~~~~~F7 & I find the Sifu platform adequate as a means to raise awareness on secure coding \\
  \hline
  ~~~~~F8 & The examples in the Sifu platform are clearly presented \\
  \hline
  ~~~~~F9 & It is fun to solve challenges in the Sifu platform \\
  \hline
  \end{tabular}
\end{table}

\section{Results}
\label{sec:results}

In this section, we present the results of the challenge feedback questions and the participants' survey.
The results were processed using RStudio version 1.2.5019.
Additionally, we briefly discuss the threats to validity.

\subsection{Challenge Feedback}
\begin{figure}[ht]
    \centering
    \includegraphics[width=.7\columnwidth]{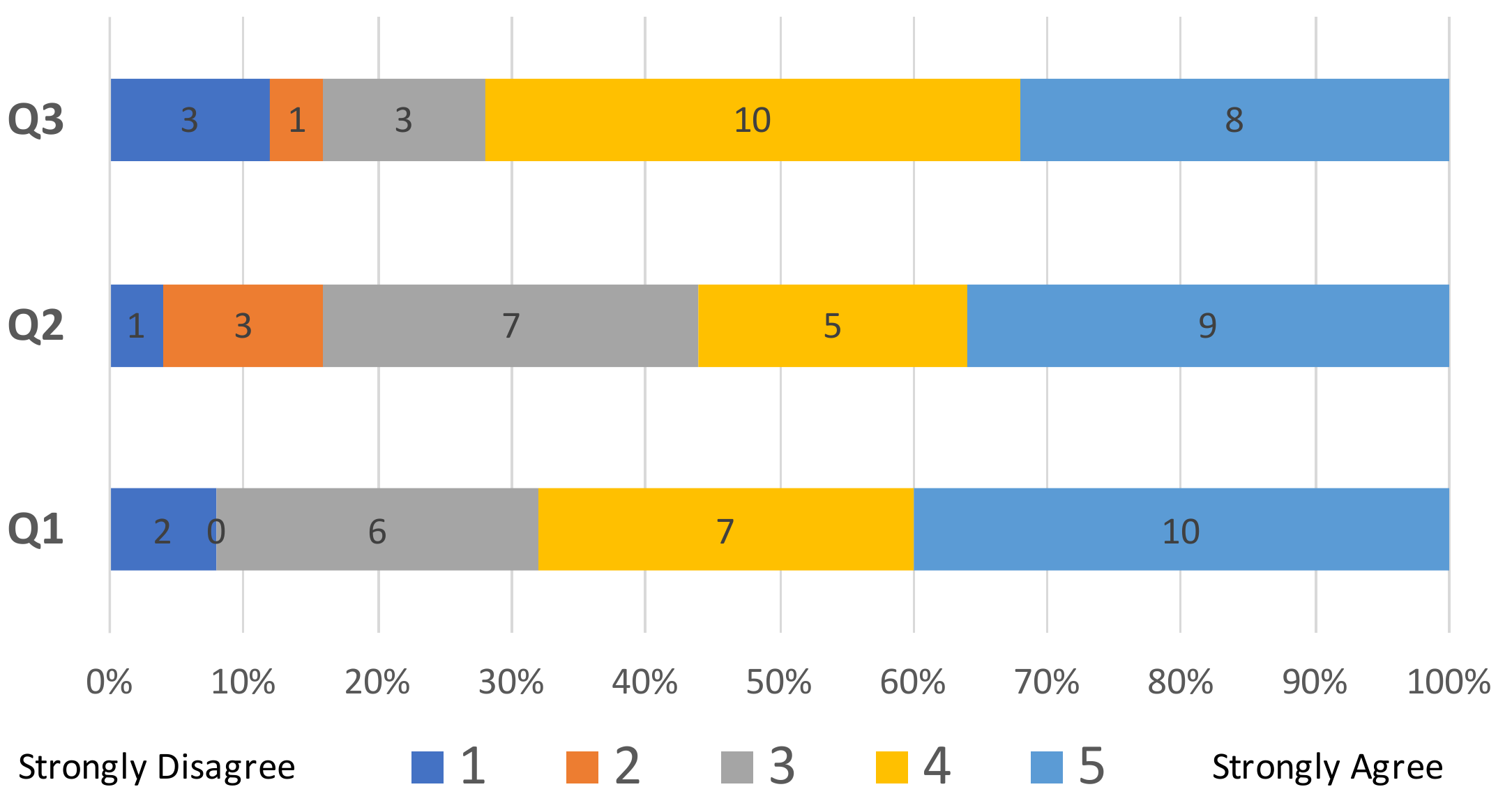}
    \caption{Evaluation of Challenges in Sifu Platform}
    \label{fig:q1q2q3}
\end{figure}

Figure~\ref{fig:q1q2q3} shows the results of the challenge rating questions.
The average values and standard deviation are the following: Q3 $3.92$ ($\sigma=1.19$), Q1 3.76 ($\sigma=1.30$), Q2 3.72 ($\sigma=1.21$).
In order of agreement: the participants are confident to be able to fix the problem in production code, have rated positively the presented challenges and would be able to recognize the vulnerability in (production) code.

\subsection{Sifu Survey}
Figure~\ref{table:platform_questions} shows the survey results.
The average values and standard deviation are the following: F6 $4.33$ ($\sigma=0.49$), F2 4.00 ($\sigma=0.38$), F9 3.93 ($\sigma=1.03$), F7 3.80 ($\sigma=0.86$), F8 3.80 ($\sigma=0.94$), F1 3.73 ($\sigma=0.70$), F3 3.67 ($\sigma=0.62$), F5 3.67 ($\sigma=1.35$), and F4 3.33 ($\sigma=0.82$).
In general, the overall positive feedback gathered through the survey shows that the Sifu platform helps to raise awareness on software developers on the topic of secure coding and secure software development best practices.
In particular, the Sifu platform helps software developers to practice secure coding guidelines (F6) and helps software developers to improve their secure coding skills (F2).
Furthermore, using the platform is fun and adequate as a means to raise secure coding awareness (F9 and F7).
In terms of awareness (perception - F3, protection - F2, and behavior - F6), as defined by Hänsch et al.~\cite{2014_Benenson_Defining_Security_Awareness}, the platform is also seen as adequate to improve awareness.
Another important aspect is that the participants find that the programming examples are clearly presented in the platform (F8).
Finally, the participants also tend to agree that using the platform can be fun (F9) and improve happiness (F5) and is an overall positive experience (F1).
The results can be split into three clusters, according to the level of agreement as follows: {\it medium agreement} ($3.33$-$3.67$), {\it higher agreement} ($3.73$-$3.8$) and {\it highest agreement} ($3.80$-$4.33$).
Using these clusters, the results can be interpreted in the following way (from highest agreement to medium agreement):

\begin{itemize}
    \item {\bf \it highest agreement}: helps to practice and improve secure coding, is fun and adequate to raise secure coding awareness
    \item {\bf \it higher agreement}: challenges are clearly presented and the experience is positive
    \item {\bf \it medium agreement}: helps to recognize vulnerable code and understand consequences and makes happy
\end{itemize}

\vspace{-.5cm}
\begin{figure}[ht]
    \centering
    \includegraphics[width=.7\columnwidth]{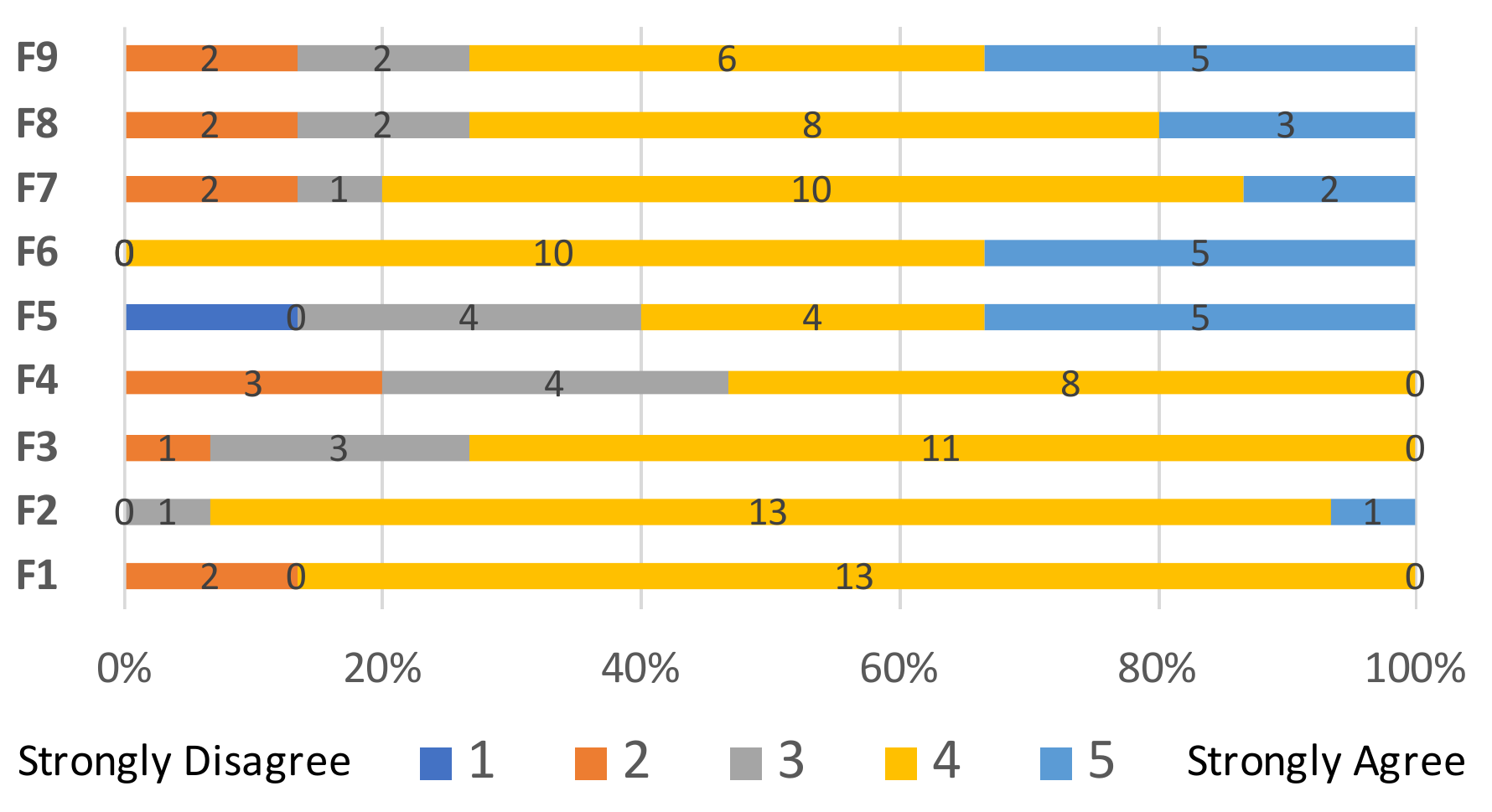}
    \caption{Survey Results}
    \label{fig:sifu_survey_results}
\end{figure}

\vspace{-.4cm}
The results hereby presented give an indication towards the suitability of the herein proposed solution to address RQ1 and RQ2, as stated in the problem statement of section \ref{sec:sifu_platform}.

%\subsection{Quotes from Participants}
%- quotes with suggestions from players
%- quotes with (positive) feedback from players

\subsection{Threats to Validity}
The main aim of this work is to present an architecture of a serious game geared towards improving the secure coding skills of software developers.
To validate the platform's usefulness, the authors have gathered feedback from 15 participants in a trial experiment.
Possible sources of threat to the validity of the results and conclusions presented in the previous section include:
\begin{itemize}
    \item {\it low number of participants}: although the gathered feedback shows a clear tendency towards positive feedback, the number of participants was low, making the standard deviations relatively high,
    \item {\it participants' background}: while the serious game is designed for industrial environments, a large portion of the participants were computer science students. Although the authors do not believe that this causes a significant change in the results, further studies with industry players is required,
    \item {\it survey design}: the survey administered at the end of the experiment was guided by survey best practices; however, it lacks a formal and thorough design, e.g., based on existing theories and existing questions database,
    \item {\it external validity}: although the goal of the present work is to propose a new method to raise secure coding awareness of software developers, our study did not contain a comparison of the methodology against existing and established methods.
\end{itemize}
\section{Conclusions}
\label{sec:conclusions}

Secure coding guidelines, secure software development best practices, and secure coding policies form an essential aspect of secure software development for industrial control and cybersystems.
Motivated by cybersecurity standards and industry needs on raising awareness about secure coding guidelines, this work presents a novel method where software developers learn these secure coding best practices in an online environment in the context of a serious game - Capture-the-Flag, while being assisted through a virtual coach.
In particular, this work addresses and details an architecture that can scale (e.g., through online training) and is based on an interview laddering technique to generate helpful hints.
Another source of inspiration for the current work is reinforcement learning techniques; however, the trainee is a human being, not a machine.

Our proposed solution uses existing open source components to perform unit-testing, static, dynamic, and run-time security analyses of the project code, which the participants need to change to eliminate software vulnerabilities.
We also briefly discuss implemented mechanisms that prevent cheating by the players and mechanisms that do not allow them to attack the system back-end.

Finally, we obtain feedback on the produced artifact through evaluation questions upon completing different challenges and a small survey at the end of the experiment.
Preliminary results show that the participants have fun using the platform and find it an adequate means to raise awareness on secure coding best practices. The developed platform will be made available in the future, after the internal software clearing process.

In future work, the authors would like to investigate additional factors that lead software developers to understand better the consequences of exploiting vulnerable code. Furthermore, the authors would like to investigate additional means to implement a more robust artificial engine for the virtual coach through systematic literature research.
Furthermore, in a future publication, the authors will perform a large-scale comparative study with existing and established cybersecurity teaching methods.
Finally, the quality of the virtual coach engine depends heavily on the quality and number of input sources.
In this aspect, the authors intend to investigate further possible sources and the quality (e.g., false positive, false negative) of the existing and future input sources.

\section*{Acknowledgements}
The authors would like to thank the participants of the survey for their time and their valuable answers.
%The authors would also like to thank Thomas Diefenbach for his helpful, insightful, and constructive comments and discussions.
This work is financed by portuguese national funds through FCT - Fundação para a Ciência e Tecnologia, I.P., under the project FCT UIDB/04466/2020. Furthermore, the third author thanks the Instituto Universitário de Lisboa and ISTAR-IUL, for their support.

% ---- Bibliography ----
\bibliographystyle{splncs04}
\bibliography{bibliography}

\begin{thebibliography}{10}
\providecommand{\url}[1]{\texttt{#1}}
\providecommand{\urlprefix}{URL }
\providecommand{\doi}[1]{https://doi.org/#1}

\bibitem{brisson2012artificial}
Brisson, A., Pereira, G., Prada, R., Paiva, A., Louchart, S., Suttie, N., Lim,
  T., Lopes, R.A., Bidarra, R., Bellotti, F., et~al.: {Artificial intelligence
  and personalization opportunities for serious games}. In: {Eighth Artificial
  Intelligence and Interactive Digital Entertainment Conference}. pp. 51--57
  (Oct 2012)

\bibitem{Davis2014}
Davis, A., Leek, T., Zhivich, M., Gwinnup, K., Leonard, W.: {The Fun and Future
  of CTF}. 2014 {USENIX} Summit on Gaming, Games, and Gamification in Security
  Education (3GSE 14) pp.~1--9 (2014),
  \url{https://www.usenix.org/conference/3gse14/summit-program/presentation/davis}

\bibitem{dobrovsky2016approach}
Dobrovsky, A., Borghoff, U.M., Hofmann, M.: {An approach to interactive deep
  reinforcement learning for serious games}. In: {2016 7th IEEE International
  Conference on Cognitive Infocommunications (CogInfoCom)}. pp. 85--90. {IEEE}
  (2016)

\bibitem{2016_Doerner_Serious_Games}
Dörner, R., Göbel, S., Effelsberg, W., Wiemeyer, J.: {Serious Games:
  Foundations, Concepts and Practice}. Springer International Publishing, 1
  edn. (2016). \doi{10.1007/978-3-319-40612-1}

\bibitem{Maria2019}
{Frey}, S., {Rashid}, A., {Anthonysamy}, P., {Pinto-Albuquerque}, M., {Naqvi},
  S.A.: {The Good, the Bad and the Ugly: A Study of Security Decisions in a
  Cyber-Physical Systems Game}. IEEE Transactions on Software Engineering
  \textbf{45}(5),  521--536 (2019)

\bibitem{gasiba_re19}
Gasiba, T., Beckers, K., Suppan, S., Rezabek, F.: {On the Requirements for
  Serious Games geared towards Software Developers in the Industry}. In:
  Damian, D.E., Perini, A., Lee, S. (eds.) 27th {IEEE} International
  Requirements Engineering Conference, {RE} 2019, Jeju Island, Korea (South),
  September 23-27, 2019. IEEE (2019),
  \url{https://ieeexplore.ieee.org/xpl/conhome/8910334/proceeding}

\bibitem{gasiba2020_ranking_scg}
Gasiba, T., Lechner, U., Cuellar, J., Zouitni, A.: {Ranking Secure Coding
  Guidelines for Software Developer Awareness Training in the Industry} (Jun
  2020), {Accepted for Publication}

\bibitem{gasiba2020_challenge_types}
Gasiba, T., Lechner, U., Pinto-Albuquerque, M., Zouitni, A.: {Design of Secure
  Coding Challenges for Cybersecurity Education in the Industry} (Sep 2020),
  {Accepted for Publication}

\bibitem{2018_Graziotin_Happy_Developers}
Graziotin, D., Fagerholm, F., Wang, X., Abrahamsson, P.: What happens when
  software developers are (un)happy. Journal of Systems and Software
  \textbf{140},  32--47 (Jun 2018)

\bibitem{Groves2009}
Groves, R.M., Fowler, F., Couper, M., Lepkowski, J., Singer, E.: {Survey
  Methodology}. John Wiley \& Sons, 2 edn. (2009)

\bibitem{2014_Benenson_Defining_Security_Awareness}
Hänsch, N., Benenson, Z.: Specifying {IT} security awareness. In: 25th
  International Workshop on Database and Expert Systems Applications, Munich,
  Germany. pp. 326--330 (Sep 2014). \doi{10.1109/DEXA.2014.71}

\bibitem{2018_62443_4_1}
{IEC 62443-4-1}: Security for industrial automation and control systems - part
  4-1: Secure product development lifecycle requirements. Standard,
  {International Electrotechnical Commission} (Jan 2018)

\bibitem{2013_27001}
{ISO 27001}: {Information technology -- Security techniques -- Information
  security management systems -- Requirements}. Standard, {International
  Standard Organization}, Geneva, CH (Oct 2013)

\bibitem{gitlab_2019}
Patel, S.: {2019 Global Developer Report: DevSecOps finds security roadblocks
  divide teams} (July 2020),
  \url{https://about.gitlab.com/blog/2019/07/15/global-developer-report/},
  [Online; posted on July 15, 2019]

\bibitem{2018_rieb_IT_sicherheit}
Rieb, A.: {IT-Sicherheit: Cyberabwehr mit hohem Spaßfaktor}. In: kma - Das
  Gesundheitswirtschaftsmagazin. vol.~23, pp. 66--69 (Jul 2018)

\bibitem{2017_rieb_gamified_approach}
Rieb, A., Gurschler, T., Lechner, U.: A gamified approach to explore techniques
  of neutralization of threat actors in cybercrime. In: {GDPR \& ePrivacy}: APF
  2017 - Proceedings of the 5th ENISA Annual Privacy Forum. pp. 87--103.
  Lecture Notes in Computer Science, Springer Verlag (Jun 2017)

\bibitem{rietz2019ladderbot}
Rietz, T., Maedche, A.: {LadderBot: A Requirements Self-Elicitation System}.
  In: 2019 IEEE 27th International Requirements Engineering Conference (RE).
  pp. 357--362. IEEE (2019)

\bibitem{schneier_2019_sw_devel}
Schneier, B.: {Software Developers and Security} (July 2020),
  \url{https://www.schneier.com/blog/archives/2019/07/software\_develo.html},
  {Online}

\bibitem{siemens02_charter}
{Siemens AG}: {Charter of Trust} (July 2020),
  \url{https://www.charteroftrust.com/}, {Online}

\bibitem{simoes_icpec2020}
Simões, A., Queirós, R.: {On the Nature of Programming Exercises}. In: {ICPEC
  - First International Computer Programming Education Conference}. vol.~81,
  pp. 251--259. {Virtual Conference} (Jun 2020)

\bibitem{vasconcelos_icpec2020}
Vasconcelos, P., Ribeiro, R.P.: {Using Property-Based Testing to Generate
  Feedback for C Programming Exercises}. In: {ICPEC - First International
  Computer Programming Education Conference}. vol.~81, pp. 285--294. {Virtual
  Conference} (Jun 2020)

\bibitem{votipka2018toward}
Votipka, D., Mazurek, M.L., Hu, H., Eastes, B.: {Toward a Field Study on the
  Impact of Hacking Competitions on Secure Development}. In: {Workshop on
  Security Information Workers (WSIW)}. {Marriott Waterfront - Baltimore, MD,
  USA} (Aug 2018)

\bibitem{WhiteSource2019}
WhiteSource: {What are the Most Secure Programming Languages?} (Mar 2019),
  \url{https://www.whitesourcesoftware.com/most-secure-programming-languages/}

\end{thebibliography}

\end{document}